\documentclass[epj]{svjour}

\usepackage{graphicx}
\def\be{\begin{equation}}
\def\ee{\end{equation}}
\def\bea{\begin{eqnarray}}
\def\eea{\end{eqnarray}}
\def\nnb{\nonumber}
\def\Rate{{\cal R}}

\def\log{\ln}

\newcommand{\e}{\epsilon}

\headnote{\hfill \parbox{4cm}{\vspace{-2cm}
\tt \normalsize CERN-TH/2003-220}}

\title{Resummed jet rates for heavy quark production \\ in e$^+$e$^-$ annihilation
\thanks{Work supported by the EC 5th Framework Programme under
        contract numbers HPMF-CT-2000-00989 and HPMF-CT-2002-01663.}}

\author{
Germ\'an Rodrigo~\thanks{\email{German.Rodrigo@cern.ch}}
\and
Frank Krauss~\thanks{\email{Frank.Krauss@cern.ch}. Present address: Institut 
f\"ur Theoretische Physik, TU Dresden, D-01062 Dresden, Germany.}}

\institute{
Theory Division, CERN, CH-1211 Geneva 23, Switzerland.}

\date{Received: September 30, 2003}

\abstract{Expressions for Sudakov form factors for heavy quarks are presented. 
They are used to construct resummed jet rates in $e^+e^-$ annihilation. 
Predictions are given for production of bottom quarks at LEP 
and top quarks at the Linear Collider.}


\begin{document}

%%%%%%%%%%%%%%%%%%%%%%%%%%%%%%%%%%%%%%%%%%%%%%%%%%%%%%%%%%%%%%%%%
%\authorrunning{}
%\titlerunning{}

\maketitle

%%%%%%%%%%%%%%%%%%%%%%%%%%%%%%%%%%%%%%%%%%%%%%%%%%%%%%%%%%%%%%%%%

\section{Introduction}

The formation of jets is the most prominent feature of perturbative QCD in 
$e^+e^-$ annihilation into hadrons. Jets can be visualized as large
portions of hadronic energy or, equivalently, as a set of hadrons confined 
to an angular region in the detector. In the past, this qualitative 
definition was replaced by quantitatively precise schemes to define and 
measure jets, such as the cone algorithms of the Weinberg--Sterman 
\cite{Sterman:1977wj} type or clustering algorithms, e.g. the Jade 
\cite{Bartel:1986ua,Bethke:1988zc} or the Durham scheme ($k_\perp$ scheme) 
\cite{Catani:1991hj}. A refinement of the latter one is provided by the 
Cambridge algorithm \cite{Dokshitzer:1997in}.  
Equipped with a precise jet definition the determination of jet production 
cross sections and their intrinsic properties is one of the traditional 
tools to investigate the structure of the strong interaction and to deduce 
its fundamental parameters. In the past decade, precision measurements, 
especially in $e^+e^-$  annihilation, have established both the gauge 
group structure underlying QCD 
and the running of its coupling constant $\alpha_s$ over a wide range of 
scales. In a similar way, also the quark masses should vary with the scale. 

A typical strategy to determine the mass of, say, the bottom-quark at the 
centre-of-mass (c.m.) energy of the collider is to compare the ratio of 
three-jet production cross sections for heavy and light quarks 
\cite{Rodrigo:1997gy,Abreu:1997ey,Brandenburg:1999nb,Barate:2000ab,Abbiendi:2001tw}. 
At jet resolution scales below the mass of the quark, i.e. for gluons 
emitted by the quark with a relative transverse momentum $k_\perp$ smaller 
than the mass, the collinear divergences are regularized by the quark mass. 
In this region mass effects are enhanced by large logarithms $\log(m_b/k_\perp)$, 
increasing the significance of the measurement. Indeed, this leads to a 
multiscale problem since in this kinematical region also large logarithms 
$\log(\sqrt{s}/k_\perp)$ appear such that both logarithms need to be resummed 
simultaneously.  
A solution to a somewhat similar two-scale problem, namely for 
the average sub-jet multiplicities in two- and three-jet events in $e^+e^-$ 
annihilation was given in~\cite{Catani:1992tm}.
We report here on the resummation of such logarithms
in the $k_\perp$-like jet algorithms~\cite{Krauss:2003cr} and provide some 
predictions for heavy quark production. A preliminary comparison with 
next-to-leading order calculations of the three-jet 
rate~\cite{Rodrigo:1997gy,Rodrigo:1996ha,Bilenky:1998nk,Rodrigo:1999qg}  
is presented.

\section{Jet rates for heavy quarks}

A clustering according to the relative transverse momenta has a number of 
properties that minimize the effect of hadronization corrections and allow 
an exponentiation of leading (LL) and next-to-leading logarithms 
(NLL)~\cite{Catani:1991hj} stemming from soft and collinear 
emission of secondary partons. Jet rates in $k_\perp$ algorithms can be 
expressed, up to NLL accuracy, via integrated splitting 
functions and Sudakov form factors~\cite{Catani:1991hj}. 
For a better description of the jet properties, however, the matching 
with fixed order calculations is mandatory.
Such a matching procedure was first defined for event shapes in~\cite{Catani:1992ua}.
Later applications include the matching of fixed-order and resummed
expressions for the four-jet rate in $e^+e^-$ annihilation into massless 
quarks~\cite{Dixon:1997th,Nagy:1998bb}. A similar scheme for the matching
of tree-level matrix elements with resummed expressions in the framework
of Monte Carlo event generators for $e^+e^-$ processes was suggested 
in~\cite{Catani:2001cc} and extended to general collision types 
in~\cite{Krauss:2002up}. 

We shall recall here the results obtained in~\cite{Krauss:2003cr} for 
heavy quark production in $e^+e^-$ annihilation.
In the quasi-collinear limit~\cite{Catani:2000ef,Catani:2002hc}, the squared amplitude at 
tree-level fulfils a factorization formula, where the splitting functions $P_{ab}$ for the 
branching processes $a\to b+c$, with at least one of the partons being a heavy quark,
are given in $D=4-2\e$ dimensions by 
\begin{eqnarray}
P_{QQ}(z,q) &=& C_F \left[ \frac{1+z^2}{1-z}-\e \ (1-z) - 
                   \frac{2z(1-z)m^2}{q^2+(1-z)^2 m^2}\right]~, \nnb \\
P_{gQ}(z,q) &=& T_R \left[ 1 - \frac{2z(1-z)}{1-\e} +
                      \frac{2z(1-z)m^2}{(1-\e)(q^2+m^2)} \right]~, 
\label{eq:PQQ}
\end{eqnarray}
where $z$ is the usual energy fraction of the branching, and 
$q^2$ is the space-like transverse momentum. 
As expected, these splitting functions match the massless splitting 
functions in the limit $m\to 0$ for $q^2$ fixed. Furthermore, 
for $q \ll (1-z) \  m$ the splitting function $P_{QQ}$ is not any more 
enhanced at $z\to 1$, which is the known ``dead cone''~\cite{Dokshitzer:fd}
effect. The splitting function 
\begin{eqnarray}
P_{gg}(z) &=& C_A \left[ \frac{z}{1-z} + \frac{1-z}{z}
                    + z(1-z) \right]~.
\end{eqnarray}
obviously does not get mass corrections at the lowest order. 

Branching probabilities are defined through~\cite{Krauss:2003cr}
\bea\label{full}
& & 
\Gamma_Q(Q,q,m) = \int\limits_{q/Q}^{1-q/Q} dz 
\frac{q^2}{q^2+(1-z)^2m^2} \,P_{QQ}(z,q) \nnb \\
& & \quad = \Gamma_Q(Q,q,m=0) + 
C_F\Bigg[\frac12 - \frac{q}{m} \arctan \left(\frac{m}{q}\right) \nnb \\ 
& & \quad -\frac{2m^2-q^2}{2m^2} \log \left(\frac{m^2+q^2}{q^2}\right)
\Bigg]~,  \nnb \\           
& & \Gamma_f(Q,q,m) = \int\limits_{q/Q}^{1-q/Q} dz 
\frac{q^2}{q^2+m^2} \,P_{gQ}(z,q) \nnb \\
& & \quad = T_R\,\frac{q^2}{q^2+m^2}\, \left[1 - \frac13\frac{q^2}{q^2+m^2}\right]~, \nnb \\  
& & \Gamma_g(Q,q) = \int\limits_{q/Q}^{1-q/Q} dz \, P_{gg}(z)
                   = 2C_A\left(\log\frac{Q}{q}-\frac{11}{12}\right)~,
\eea
and the Sudakov form factors, which yield the probability for a parton 
experiencing no emission of a secondary parton between transverse 
momentum scales $Q$ down to $Q_0$, read
\bea\label{Suddef}
\Delta_Q(Q,Q_0) &=& \exp\left[-\int\limits_{Q_0}^Q 
                    \frac{dq}{q}\frac{\alpha_s(q)}{\pi}\Gamma_Q(Q,q)\right] 
\,,\nnb\\
\Delta_g(Q,Q_0) &=& \exp\left[-\int\limits_{Q_0}^Q 
                              \frac{dq}{q}\frac{\alpha_s(q)}{\pi}
                              \left(\Gamma_g(Q,q)+
                              \Gamma_f(Q,q)\right)\right]         
\,,\nnb\\
\Delta_f(Q,Q_0) &=& \left[\Delta_Q(Q,Q_0)\right]^2/\Delta_g(Q,Q_0)\,,
\eea
where $\Gamma_f(Q,q)$ accounts for the number $n_f^{(l,h)}$ of active light or heavy 
quarks. Jet rates in the $k_\perp$ schemes can be expressed by the former branching  
probabilities and Sudakov form factors. For the two- and the three-jet rates
\bea \label{jetrates}
& & \Rate_2 = \left[\Delta_Q(Q,Q_0)\right]^2
\,,\\
& & \Rate_3 = 2\left[\Delta_Q(Q,Q_0)\right]^2\,
            \int\limits_{Q_0}^Q\frac{dq}{q}\frac{\alpha_s(q)}{\pi}
                               \Gamma_Q(Q,q)\Delta_g(q,Q_0)
\,, \nnb 
\eea
where $Q$ is the c.m. energy of the colliding $e^+e^-$, and  $Q_0^2 = y_{\rm cut} Q^2$ 
plays the role of the jet resolution scale. Single-flavour jet rates 
in Eq.~(\ref{jetrates}) are defined from the flavour of the primary vertex.

In order to catch which kind of logarithmic corrections are resummed with 
these expressions it is illustrative to study the above formulae in the 
kinematical regime such that $Q\gg m\gg Q_0$.
Expanding in powers of $\alpha_s$, jet rates can formally be expressed as
\bea
\Rate_n = \delta_{n2} +
          \sum\limits_{k=n-2}^\infty 
          \left( \frac{\alpha_s(Q)}{\pi} \right)^k\, 
          \sum\limits_{l=0}^{2k} c^{(n)}_{kl},
\eea
where the coefficients $c^{(n)}_{kl}$ are polynomials of order $l$ in 
$L_y = \log(1/y_{\rm cut})$ and $L_m = \log(m^2/Q_0^2)$. 
The coefficients for the first order in $\alpha_s$ are given by
\bea\label{expand1}
c^{(2)}_{12} = -c^{(3)}_{12} &=& -\frac12 C_F (L_y^2-L_m^2)\,,\nnb\\
c^{(2)}_{11} = -c^{(3)}_{11} &=& \frac32 C_F L_y + \frac12 C_F L_m\,.
\label{coef1}
\eea
The coefficients for the second order in $\alpha_s$ can be found in~\cite{Krauss:2003cr},
as well as the corresponding result for the four-jet rate.

\begin{figure}
\includegraphics[width=9.5cm,height=7.8cm]{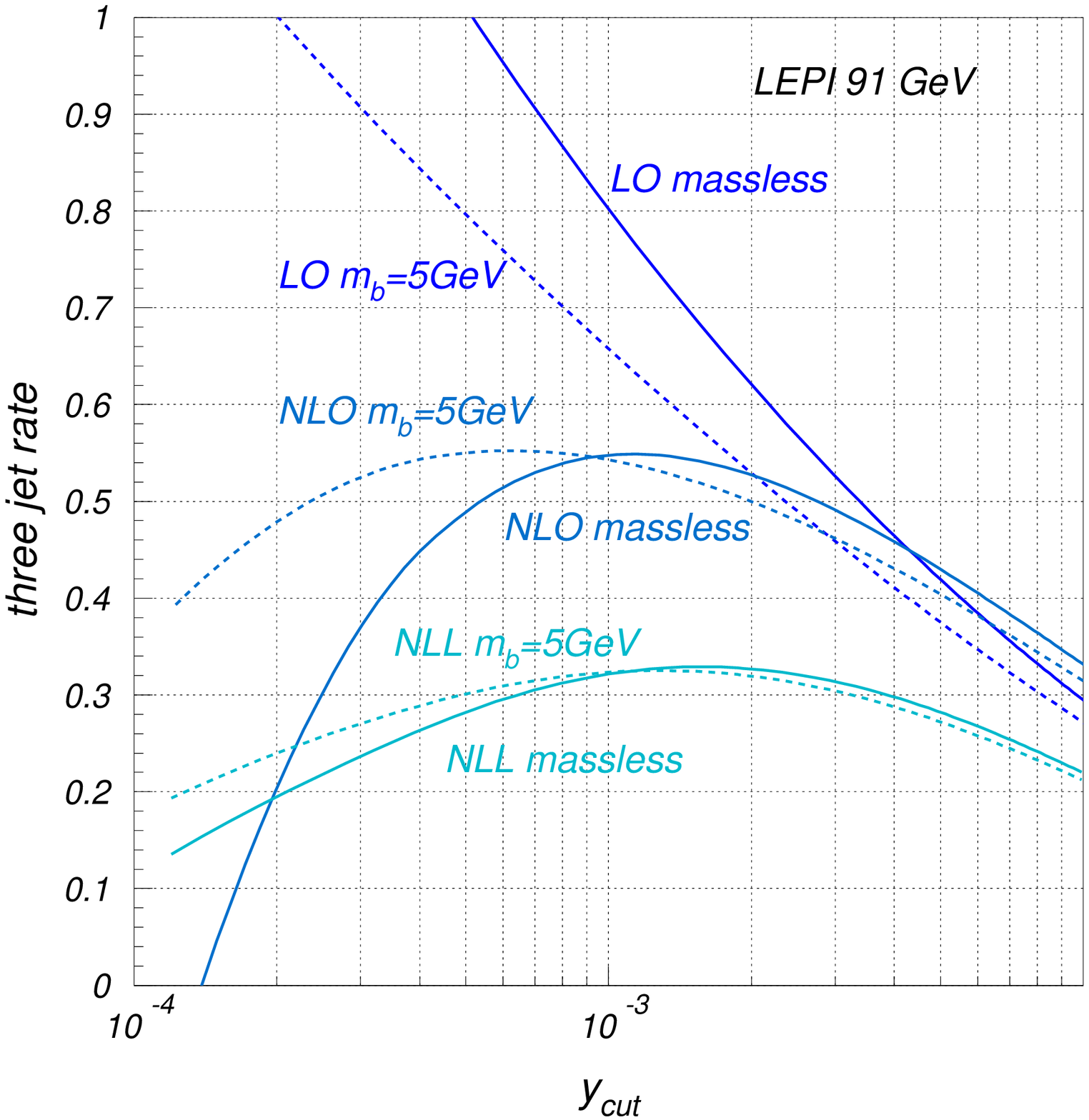} 
\includegraphics[width=9.5cm,height=7.8cm]{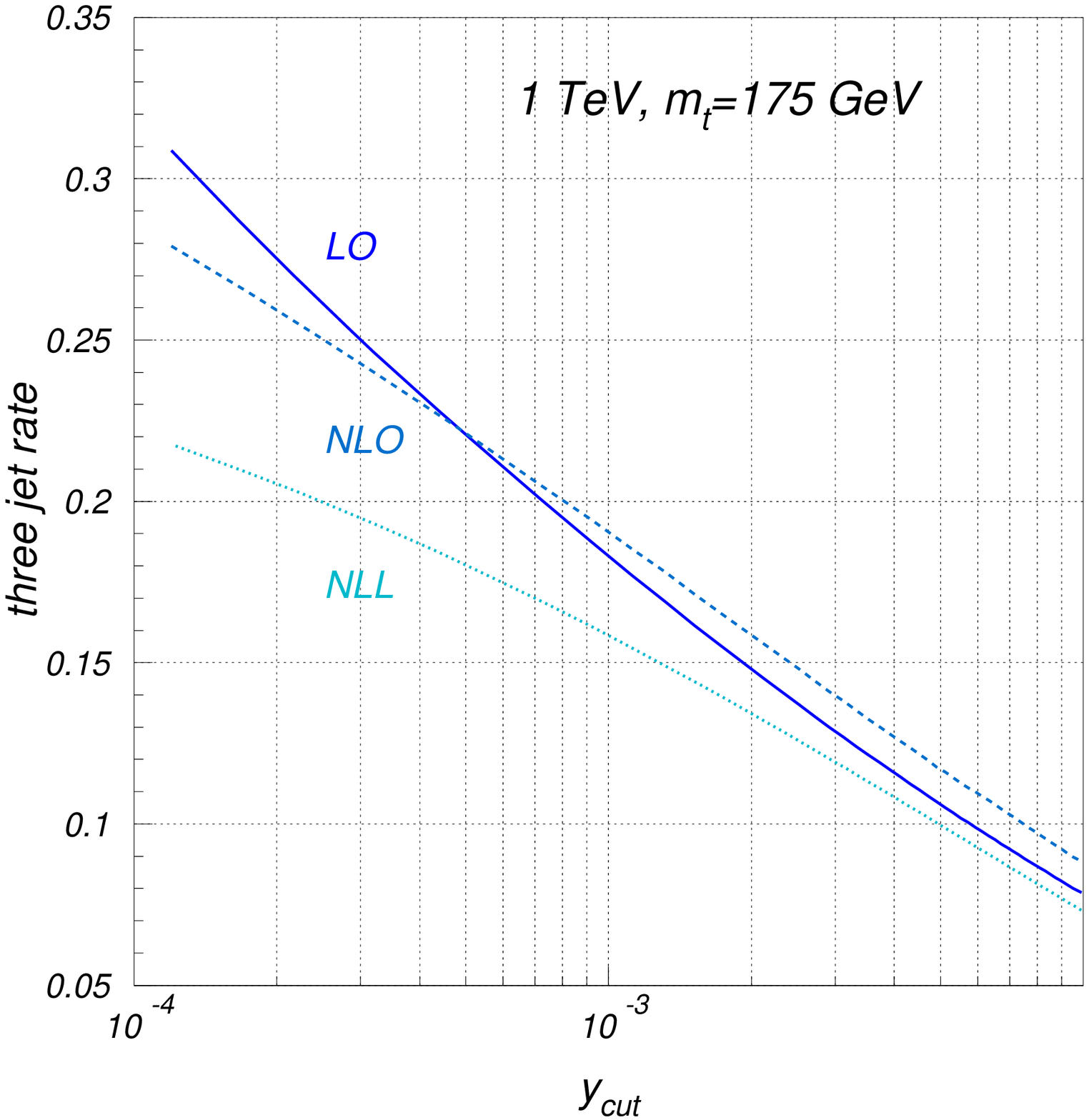} 
\caption{\label{B91} 
         Effect of a $b$-mass of $5$ GeV on the single-flavour
         three-jet rate at LEP1 energies as a function of the jet 
         resolution parameter in the Cambridge algorithm (upper plot).
         Effect of a $t$-mass of $175$ GeV on the single-flavour  
         three-jet rate as a function of the jet resolution parameter, at a 
         potential Linear Collider operating at c.m. energies of 1 TeV 
         (lower plot). Predictions are provided at LO, NLO and NLL accuracy.}
\end{figure}

The impact of mass effects can be highlighted by two examples, namely by
the effect of the bottom quark mass in $e^+e^-$ annihilation at the $Z$-pole 
(Fig.~\ref{B91}, up), and by the effect of the top quark mass at a potential 
Linear Collider operating in the TeV region (Fig.~\ref{B91}, down).
In Fig.~\ref{B91}, leading order (LO) and next-to-leading order (NLO) predictions 
for three-jet rates are compared with the NLL result as obtained by numerical 
integration from Eq.~(\ref{jetrates}). For a comparison with the dead cone 
approximation see also~\cite{Krauss:2003cr}.
While in the case of bottom quarks at LEP1 energies the overall effect of the 
quark mass is at the few-per-cent level, this effect becomes tremendous 
for top quarks at the Linear Collider. Fixed order predictions for 
$b$-quark production clearly fail at very low values of $y_\mathrm{cut}$,
by giving unphysical values for the jet rate, while the NLL predictions
keep physical and have the correct shape. The latter is an indication 
of the necessity for performing such kind of resummations. 
Fixed order predictions work well for top production at the Linear Collider,
a consequence of the strong cancellation of leading logarithmic corrections,
and are fully compatible with our NLL result.

\section{Conclusions}

Sudakov form factors involving heavy quarks have been 
employed to estimate the size of mass effects in jet rates in 
$e^+e^-$ annihilation into hadrons. These effects are sizeable and therefore 
observable in the experimentally relevant region. 
A preliminary comparison with fixed order results have been presented,
and showed good agreement.  
Matching between fixed-order calculations and resummed results
is in progress~\cite{inprogress}.

\section*{Acknowledgements}

We would like to thank Z.~Trocsanyi for his kind invitation 
to present these results at the HEP2003 Europhysics Conference in Aachen.
G.R. acknowledges partial support from 
Generalitat Valenciana under grant CTIDIB/ 2002/24 and 
MCyT under grants FPA-2001-3031 and BFM 2002-00568.

\end{document}